\documentclass[pre,twocolumn,aps,floatfix,superscriptaddress]{revtex4}

\usepackage[english]{babel}
\usepackage{amssymb,amsfonts,amsmath}
\usepackage[pdftex]{graphicx}
\usepackage{latexsym}
\usepackage{multirow}
\usepackage{color}
\usepackage{ltxtable}

\usepackage{amsmath}
\usepackage{hyperref}

\begin{document}


\title{Does network complexity help organize Babel's library?}

\author{Juan Pablo C\'ardenas}
\affiliation{Net-Works, Angamos 451. Vi\~na del Mar,  Chile.}

\author{Iv\'an Gonz\'alez}
\affiliation{Grupo de Sistemas Complejos, Dept. F\'isica y Mec\'anica Fundamentales aplicadas a la I. Agroforestal, E.T.S.I. Agr\'onomos. Universidad Polit\'ecnica de Madrid (UPM). Ciudad Universitaria, s/n, 28040, Madrid, Spain }
\author{Gerardo Vidal}
\affiliation{Pontificia Universidad Cat\'olica de Valpara\'iso. Centro de Estudios y Asistencia Legislativa-CEAL. Avenida Brasil, 2950, Valpara\'iso, Chile. }
\author{Miguel Fuentes}
\affiliation{Santa Fe Institute, 1399 Hyde Park Road, Santa Fe, New Mexico 87501, USA}
\affiliation{Instituto de Investigaciones Filos\'oficas, Buenos Aires 1428, Argentina}
\affiliation{Universidad San Sebasti\'an, Lota 246, Santiago 7500000, Chile}
\begin{abstract}

Quantitative linguistics has contributed significantly to the study of the origins and properties of natural languages. Among many statistical properties described, one of the most basic but no less fundamental, is the well-known Zipf's law. Its ubiquity would reveal underlying principles of natural languages functioning. Although the presence of this law reduces drastically the probability that a text can be the result of a random process, it is insufficient information to determine if the text is the result of a ciphering process that shifts the position of words of a ``readable'' text, or if the text corresponds to a non-sense one constructed with a ``bag of words'' with a Zipfian distribution. In this work we show that simple global topological properties of co-ocurrent word networks constructed from texts, seem to be the fingerprint of the sense texts. We observe that many statistical properties of these networks depend on the frequency of words in the text, however, others seem to be strictly determined by the grammar. Our results suggest that seems to be a \textit{lower bound of sense} that depends on the correlation between mean word connectivity and word connectivity correlation. This property, in addition to being only present in sense text, and absent in, until now, not decoded texts such as Voynich Manuscript, would also be exclusive for natural languages, allowing us to discriminate between these and formal texts.


\vspace{0.5 cm}

\textbf{keywords}: complex networks, word networks,Voynich manuscript, quantitative linguistics

\end{abstract}

\maketitle

\section{Introduction}

In 1941, Jorge Luis Borges wrote a short story about a custodian of a peculiar repository that contains all the possible arrangements of letters that can be written. In this story, the eponymous ``La Biblioteca de Babel'' (English: ``The Library of Babel'') \cite{borges} contains books of a certain size and specific characteristics. The books contain no pictures, only text; in addition, each book has 410 pages, each page has 40 rows, and each row, 80 characters. The alphabet mentioned in the story consists of only 25 orthographical symbols, including the space, the coma, and the point. Considering only these initial conditions, the library houses 25$^{1312000}$ different books.

The library thus stores all the ``sensical'' texts that could ever be written, in any language, and any of their possible variations. However, the number of sensical books is minimal in comparison to the huge number of possible ``nonsensical'' combinations of words that lack meaning in any language. The librarians of this enormous collection are both charged with its custody and obsessed with finding those books that ``say something''.

Although Borges wrote this narrative with different intentions, let us continue the Library of Babel's fictional game. In order to find those texts with meaning, some reported statistical properties of sensical texts might be used as a first filter  \cite{montemurro1}. One of the most well-known and basic, but no less important, properties comes from Zipf's law \cite{zipf}. Roughly speaking, this law says that the number of times a word appears in a text is a function of its ranked frequency of occurrence. The apparent ubiquitious applicability of the law to natural languages would reveal languages functioning \cite{Fer3}. The problem is that a text that follows Zipf's law does not necessarily have to make sense. For illustration, in the library of Babel there is a book called the ``\textit{Voynich manuscript}''  that fits the law \cite{landini} and other properties of ``real'' books \cite{montemurro}, but has been unreadable and illegible for centuries. The manuscript can not be classified either as a hoax \cite{rugg,schinner} or a text with some sort of sophisticated encryption. The reason is that a seemingly nonsensical or encrypted text such as this manuscript could actually be a readable, but scrambled, text (keeping the frequency of words, but changing their order).  Thus, although the text makes no sense for a reader who reads it word by word, the co-occurrence of the words would still comply with Zipf's law.

Another problem in the Library would be the existence of texts that only make sense to certain machines. These texts were written for machines using formal programing languages. Formal languages correspond to the set of strings of symbols and may be constrained by grammatical rules. Their alphabets, frequently required to be finite \cite{reghizzi}, are the set of symbols, letters, or tokens from which the strings of the language may be formed. Strings formed from this alphabet can be found as words in texts present in the Library. Although formal languages have been developed intentionally and their ``words'' 
are linked by grammar specific to each
, they appear very similar to those words linked in a text written in a natural language, making identification as strictly ``machine codes'' very difficult.

Problems such as those previously mentioned was precisely what motivated this work: searching for statistical properties that are unique to sensical texts. Using techniques borrowed from the field of complex networks, this work looks for the topological properties of co-occurrent word networks that depend exclusively on the sense of a text.  These must capture the use of a certain grammar that would be absent in senseless or ciphered texts. Thus, by using corpuses written in different natural and formal languages, we studied the networks that represent those texts in order to obtain their common properties and to detect which of these properties might allow the librarians of Babel to organize their books. 

This paper is structured as follows. In the next section, we present the method for constructing word networks and the one used to cipher them. In Section 3, we present the results of the word networks analysis. In the last section, we discuss the major implications of the obtained results and present the conclusions of our work.

\section{Network construction}

A text can be represented as a graph $G(W,E)$, where $W$ is the set of nodes, corresponding to different words contained in the text, and $E$, the set of undirected edges between them. In this work, we define an edge as the link that joins two co-occurrent (adjacent) words in the text.

\begin{figure}[h]
  \centering
  \includegraphics[width=0.35\textwidth]{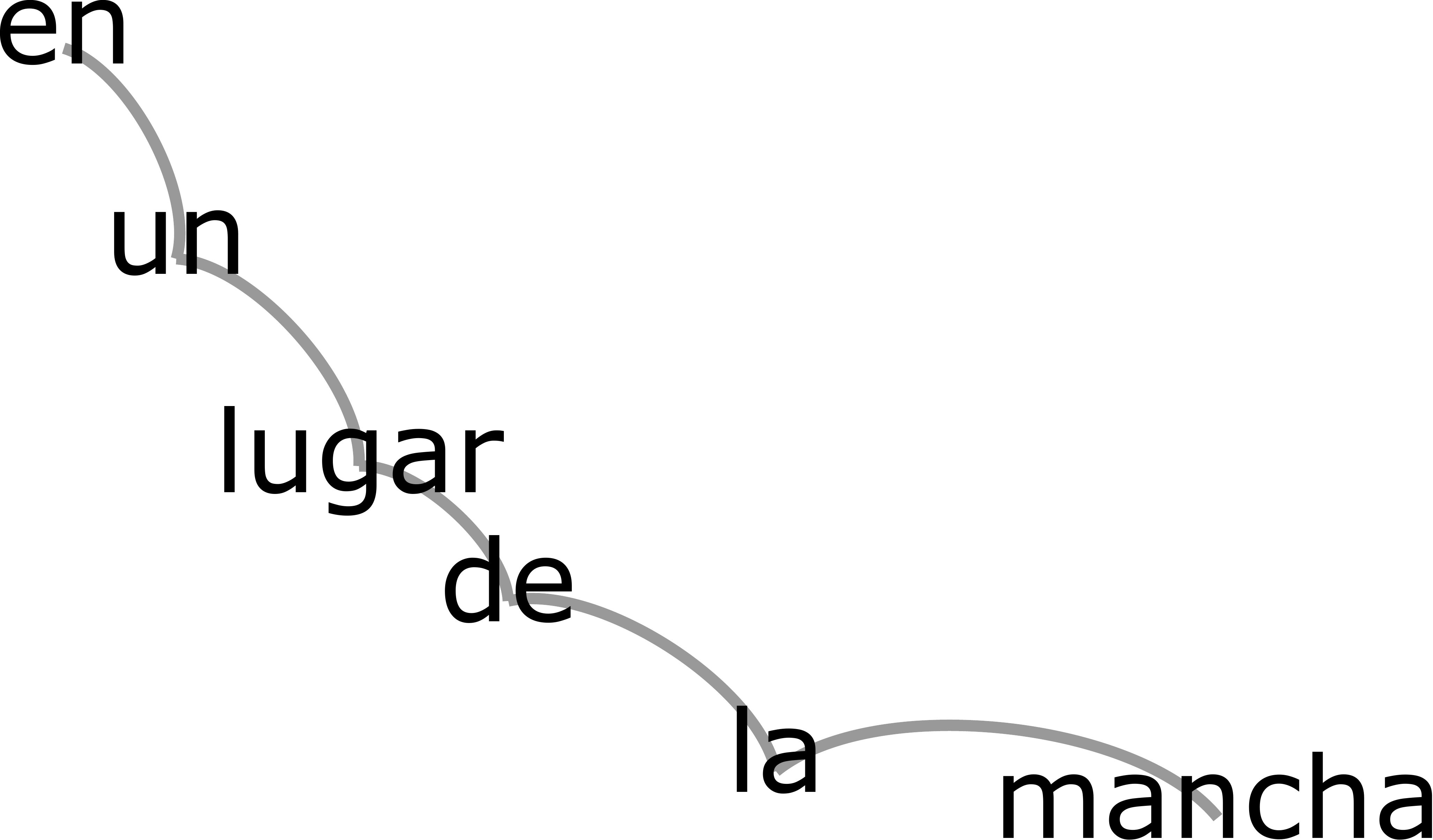}
  \caption{Network extracted from the phrase: \textit{``En un lugar de La Mancha.''}.}
  \label{fig:net}
\end{figure}

\begin{figure}[h]
  \centering
  \includegraphics[width=0.45\textwidth]{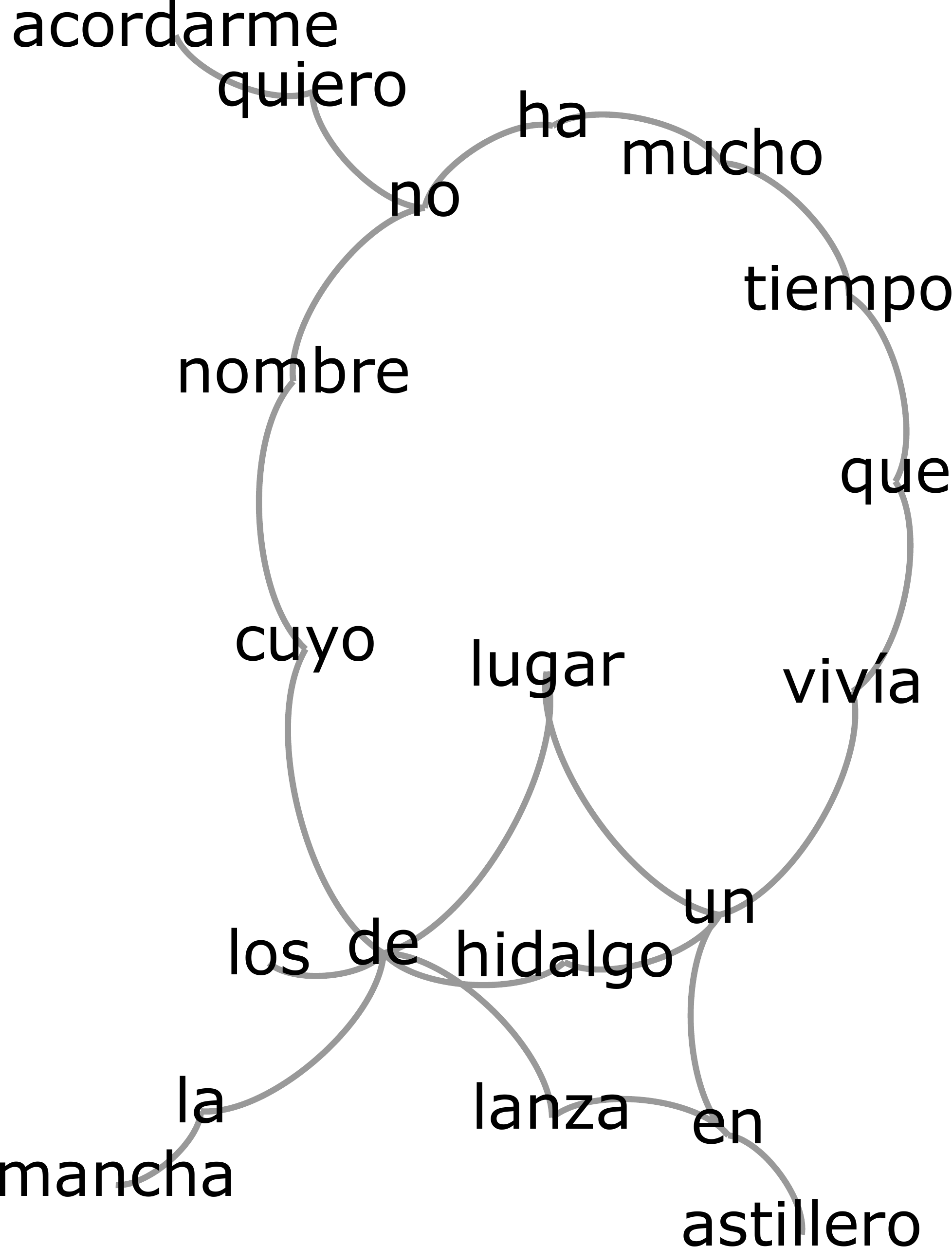}
  \caption{Network of the text:  \textit{``En un lugar de La Mancha, de cuyo nombre no quiero acordarme, no ha mucho tiempo que viv\'ia un hidalgo de los de lanza en astillero.''}.}
  \label{fig:net2}
\end{figure}

Figure \ref{fig:net} shows a simple word network constructed from a single sentence, extracted from the masterpiece of Miguel de Cervantes y Saavedra, ``El ingenioso hidalgo don Quijote de la Mancha''. The sentence,  \textit{``En un lugar de La Mancha,''}, contains six different words. Word $w_3$ =``\textit{lugar}'' has as input word $w_2$=``\textit{un}'', and as output word $w_4$=``\textit{de}''. Now, in the created graph $G$, the connectivity of $w_3$ is $k_3$=2. If the text were composed only of different words, as in this example, the network would be a simple lineal chain of $W$ different words and $W-1$ links. In this type of network, except for the words located at the beginning and at the end of the text, all the words have connectivity $k_i=2$. Nevertheless, this scenario is quite improbable in long texts because in natural languages, as mentioned above, words are used with different frequency. Thus, if we introduce a new sentence to the previous text, for example ``\textit{de cuyo nombre no quiero acordarme, no ha mucho tiempo que viv\'ia un hidalgo de los de lanza en astillero}'', there are both new and repeated words added. In our process of network generation, repeated words maintain the same number $w_{i}$ that corresponds to its first appearance in the text; however, that word can be subsequently connected to different words or connected many times to the same word (imagine a character name, composed of two words, that appears many times throughout the text, \textit{e.g.}, Don Quijote).  The network generated after the incorporation of the new sentence is shown in Figure \ref{fig:net2}. Notice that the words ``\textit{de}'', ``\textit{no}'', ``\textit{en}'' and ``\textit{un}'' have a higher connectivity than the rest of the words.

It must be pointed out that the method proposed in this work is case-insensitive, eliminating capitalization effects. Furthermore,  this work assumes that (any) punctuation cuts any relationship between two words. For this reason, in the example above, the words ``\textit{mancha}'' and ``\textit{de}'', and ``\textit{acordarme}'' and ``\textit{no}'', are not connected. 

\subsection{Network ciphered}

\begin{figure}[h]
  \centering
  \includegraphics[width=0.45\textwidth]{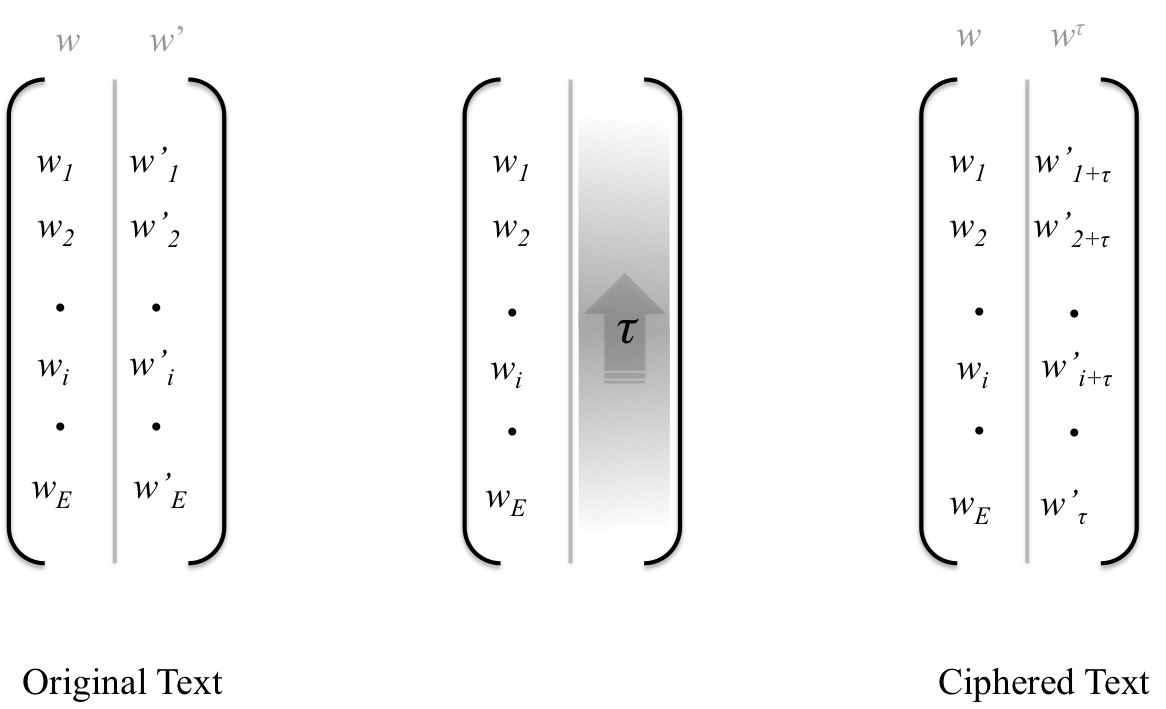}
  \caption{Process of word circular-permutation for text encryptation. }
  \label{fig:permutation}
\end{figure}

In order to study the properties of senseless texts and compare them with those with sense, we encrypt sense texts according to a simple permutation rule (Fig. \ref{fig:permutation}). Ciphered texts are obtained using the list of edges between co-ocurrent words, $ei=(w_i,w'_i)$, where $i=[1,..,E]$ and $w_i$ and $w'_i$ are the words that appear adjacent in the original text. To  encrypt the text, we maintain fixed the first column $w$ while the elements of the second column $w'$ are moved $\tau$ places, where $\tau$ is a random number. Thus, the ciphered and senseless text is composed by new edges $e_{i}$$^\tau$=($w_{i}$,$w_{i}$$^\tau$), where $w_{i}$$^\tau$ = $w'_{i+\tau}$ and $w'_{E+j}=w'_{j}$. 

\section{Results}

In order to analyze the word networks generated, we used a set of metrics to characterize the whole system as well as local relationships between words. One such basic metric used in network topology characterization is degree distribution \cite{New,Dor}, $P(k)$. This probability distribution represents, in this case, the probability of finding a word with $k$ edges in the network. Degree distribution is one of the most important characteristics of networks, especially due to the fact that the distribution of node connections is indicative of the underlying network formation mechanisms \cite{barabasi,cam}. In fact, random networks \cite{Erd}, whose graphs show randomly-chosen relationships between nodes, show homogeneous distributions of connectivity, whereas (so called) complex topologies display inhomogeneous distributions \cite{BAb}. This non-uniformity denotes the presence of rules (\textit{e.g.}, grammar rules in word networks) or mechanisms that distribute the links unequally, and where some few nodes may concentrate the bulk of connections. 

\begin{figure}[h]
	\centering
		\includegraphics [width=0.32\textwidth] {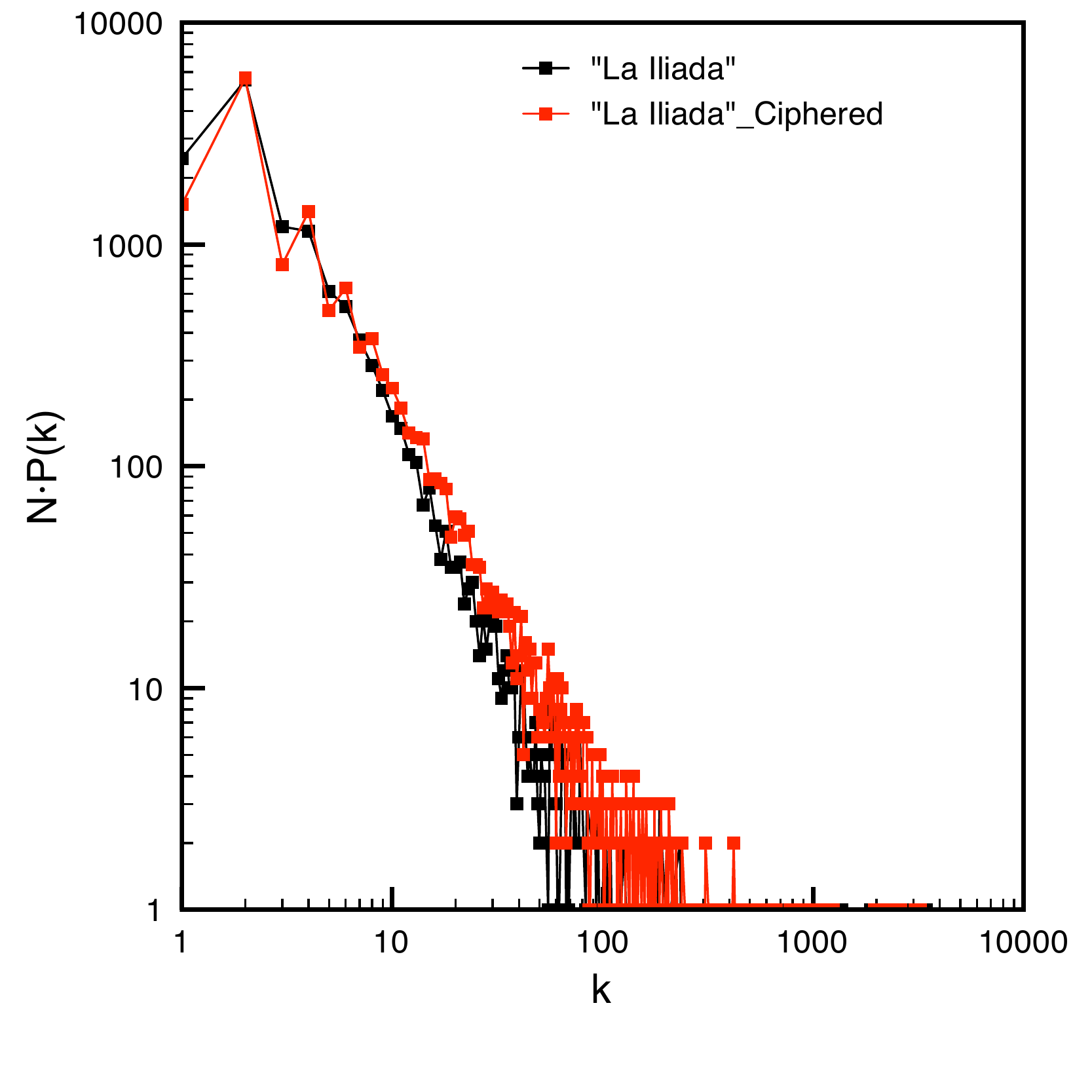}
		 \includegraphics [width=0.32\textwidth] {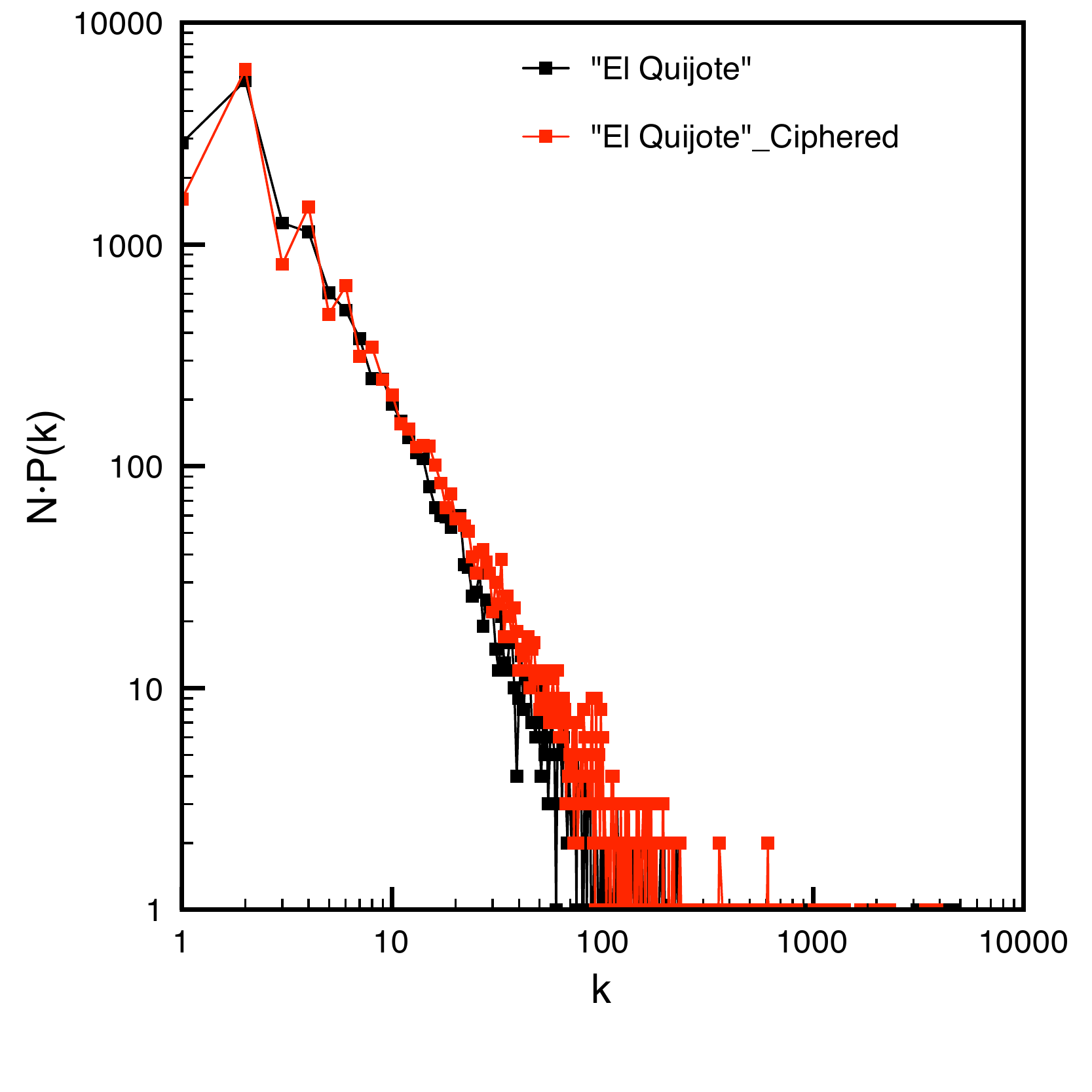}
		 \includegraphics [width=0.32\textwidth] {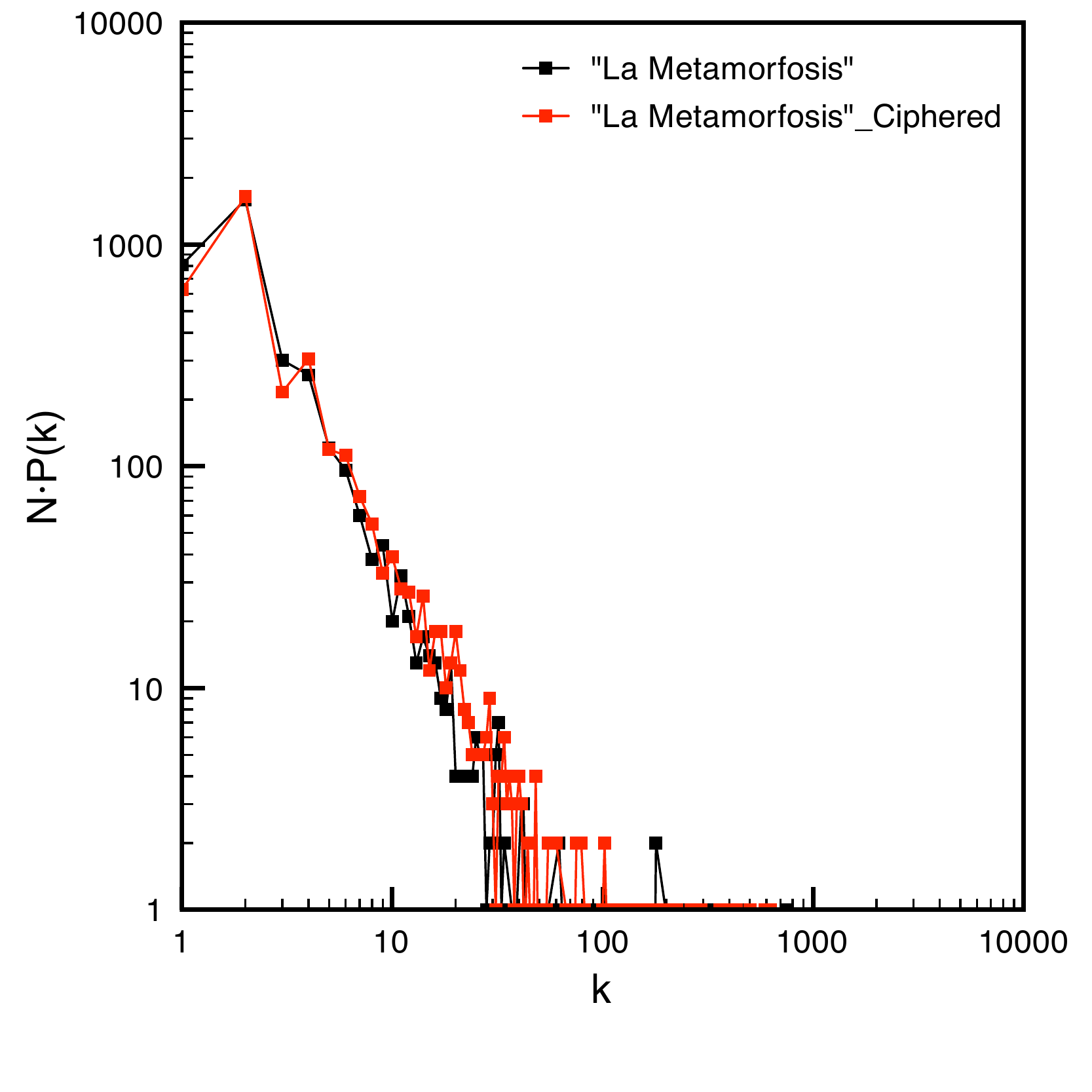}
\caption{Distribution of the number of words with $k$ connections, $N \cdot P(k)$, for the texts. Top: ``\textit{La Iliada}'' ($W = 3895$)
by Homer. Mid: ``\textit{El ingenioso hidalgo don Quijote de la Mancha}'' ($W = 3895$) by Miguel de Cervantes y Saavedra. Bottom: ``\textit{La Metamorfosis}'' by Franz Kafka ($W = 3895$). Original text in black squares and ciphered version in red squares.} 
	\label{fig:pk}
\end{figure}

In fact, this is what we found in the word networks studied. Figure \ref{fig:pk} shows this behavior consistent with complex topologies, as the distribution of the number of words with $k$ connections, $N \! \cdot \! P(k)$, for three classic books: \textit{``La Iliada''}, \textit{``El ingenioso hidalgo don Quijote de la Mancha''} and \textit{``La Metamorfosis''}. As can be seen, the distributions show word networks with a wide range of connectivities: many words are poorly connected, while a few are densely connected. Moreover, the log-log plots of the distributions tend towards a straight line, which indicates that the distributions of the three books follow a power-law-like form, $N \!\cdot \!P(k) \sim k^{-\gamma}$, where $\gamma$ is the scaling exponent. This scaling, typically observed in complex systems \cite{Newb}, reveals a high level of inhomogeneity in the number of connections among words in the network. Thus, in networks constructed from natural languages, like the Spanish presented in these examples, most of the words have few connections, while there are a handful of words responsible for a large majority of connections (\textit{hubs}).  These hubs play an important role \cite{sole} and correspond to articles and conjunctions mainly. 

When we analyzed the same texts but ciphered, we observed that the distributions are similar to those of the originals (red line in Figure \ref{fig:pk}), which implies that inhomogeneous distribution of connectivity depends not on the way words are linked, but rather on the frequency of words in the text. The message is clear: degree distribution in word networks gives no information on the meaning of a text, it is but a projection of Zipf's law in the case of our ciphered texts. It is necessary to emphasize that our senseless texts are different to the random texts described in \cite{Fer3}.

\begin{table}[!ht]
\caption{Topological properties of \textit{The Universal Declaration of Human Rights} written in different languages (ID=[1,17]), classic books (ID=[18,26]) and computer codes (ID=[27,36]). Number of different words $W$, number of Edges $E$, mean degree $\langle k \rangle$, mean clustering coefficient $\langle C \rangle$ and average path lenght $\langle l \rangle$ for original and ciphered texts ($\langle C_c \rangle$, $\langle l_c \rangle$).}
\centering
\begin{tiny}
\begin{tabularx}{260pt}{lX|rrcccccc} 
\hline
id&Network&$W$&$E$&$\langle k \rangle$&$\langle C \rangle$&$\langle C_c \rangle$&$\langle l \rangle$&$\langle l_c \rangle$ \\ \hline
1&Dutch&494&982&1.99&0.194&0.095$\pm$0.012&3.20&3.67$\pm$0.06 \\
2&English&430&839&1.95&0.175&0.100$\pm$0.012&3.16&3.64$\pm$0.07 \\
3&Euskera&563&772&1.37&0.040&0.031$\pm$0.008&4.41&4.92$\pm$0.15 \\
4&German&517&892&1.73&0.133&0.060$\pm$0.009&3.66&4.05$\pm$0.08 \\
5&Greek&576&1035&1.80&0.109&0.076$\pm$0.010&3.48&3.95$\pm$0.06 \\
6&Italian&512&979&1.91&0.110&0.063$\pm$0.008&3.60&3.89$\pm$0.05 \\
7&Kanuri&554&778&1.40&0.036&0.022$\pm$0.006&4.99&5.12$\pm$0.13 \\
8&Maori&342&1197&3.50&0.325&0.180$\pm$0.016&2.69&2.98$\pm$0.03 \\
9&Mapundungun&321&730&2.27&0.230&0.127$\pm$0.015&2.95&3.37$\pm$0.05 \\
10&Nahualt&521&1118&2.15&0.190&0.096$\pm$0.012&3.25&3.61$\pm$0.04 \\
11&Portugues&464&827&1.78&0.140&0.071$\pm$0.010&3.52&3.89$\pm$0.07 \\
12&Quechua&676&921&1.36&0.030&0.020$\pm$0.005&4.99&5.20$\pm$0.12 \\
13&Rumano&586&973&1.66&0.091&0.057$\pm$0.008&3.71&4.19$\pm$0.07 \\
14&Russian&601&857&1.43&0.070&0.043$\pm$0.008&4.02&4.63$\pm$0.12 \\
15&Spanish&486&915&1.88&0.183&0.090$\pm$0.010&3.29&3.76$\pm$0.07 \\
16&Tahitic&432&1015&2.35&0.437&0.188$\pm$0.018&2.66&3.26$\pm$0.05 \\
17&Zulu&562&607&1.08&0.015&0.009$\pm$0.005&7.33&7.67$\pm$0.44 \\
\hline
18&La Iliada&14004&53499&3.82&0.360&0.166$\pm$0.003&2.90&3.29$\pm$0.01 \\
19&Metamorfosis&3613&9833&2.72&0.236&0.126$\pm$0.005&3.14&3.50$\pm$0.01 \\
20&La Odisea&10982&44499&4.05&0.387&0.175$\pm$0.003&2.88&3.25$\pm$0.01 \\
21&El Quijote&14754&66961&4.54&0.446&0.194$\pm$0.002&2.81&3.20$\pm$0.01 \\
22&Harry Potter (sp)&8312&31310&3.77&0.318&0.152$\pm$0.003&2.96&3.31$\pm$0.01 \\
23&Harry Potter (en)&6024&35163&5.84&0.479&0.206$\pm$0.004&2.59&3.02$\pm$0.01 \\
24&La Biblioteca de Babel&1081&1858&1.72&0.118&0.077$\pm$0.007&3.66&4.13$\pm$0.07 \\
25&The Library of Babel&1019&2230&2.19&0.195&0.114$\pm$0.008&3.14&3.58$\pm$0.04 \\
\hline
26&Voynich manuscript&1997&8031&4.02&0.159&0.138$\pm$0.005&3.33&3.24$\pm$0.01 \\
\hline
27&C$_1$&181&582&3.22&0.226&0.174$\pm$0.019&3.47&2.82$\pm$0.03 \\
28&C$_2$&244&739&3.03&0.283&0.146$\pm$0.014&3.00&2.98$\pm$0.03 \\
29&Fortran$_1$&947&3927&4.15&0.334&0.190$\pm$0.009&2.76&2.92$\pm$0.01 \\
30&Fortran$_2$&290&732&2.52&0.190&0.098$\pm$0.012&3.00&3.23$\pm$0.04 \\
31&Fortran$_3$&2438&6908&2.83&0.200&0.115$\pm$0.005&3.25&3.43$\pm$0.01 \\
32&Fortran$_4$&221&763&3.45&0.277&0.178$\pm$0.018&2.73&2.77$\pm$0.03 \\
33&Fortran$_5$&306&899&2.94&0.200&0.138$\pm$0.015&3.07&3.03$\pm$0.03 \\
34&Fortran$_6$&337&1892&5.61&0.375&0.263$\pm$0.016&2.48&2.55$\pm$0.01 \\
35&Fortran$_7$&869&3979&4.58&0.359&0.205$\pm$0.011&2.86&2.87$\pm$0.01 \\
36&Fortran$_8$&156&443&2.84&0.270&0.18$\pm$0.02&2.83&2.86$\pm$0.05 \\
\hline

\end{tabularx} \label{tab:TABLEHR}
\end{tiny}
\end{table}

To further characterize the word networks, we computed other classic topological measures of networks: the average path length $\langle l \rangle$ between pairs of words; and the clustering coefficient of the network $\langle C \rangle$, as the average coefficient of all the words in a network. Table \ref{tab:TABLEHR} shows these metrics for different words networks constructed from \textit{The Universal Declaration of Human Rights} written in different languages (ID=[1,17]), classic books (ID=[18,25]),  the text of the ``\textit{Voynich manuscript}'' (ID=26), and a set of computer codes written with formal grammatical rules in two different programing languages \footnote{Formal languages are used as the basis for defining the grammar of programming languages and formalized versions of subsets of natural languages in which the words of the language represent concepts that are associated with particular meanings or semantics.} (ID=[27,36]). 

\begin{figure}[h]
	\centering
		\includegraphics [width=0.47\textwidth] {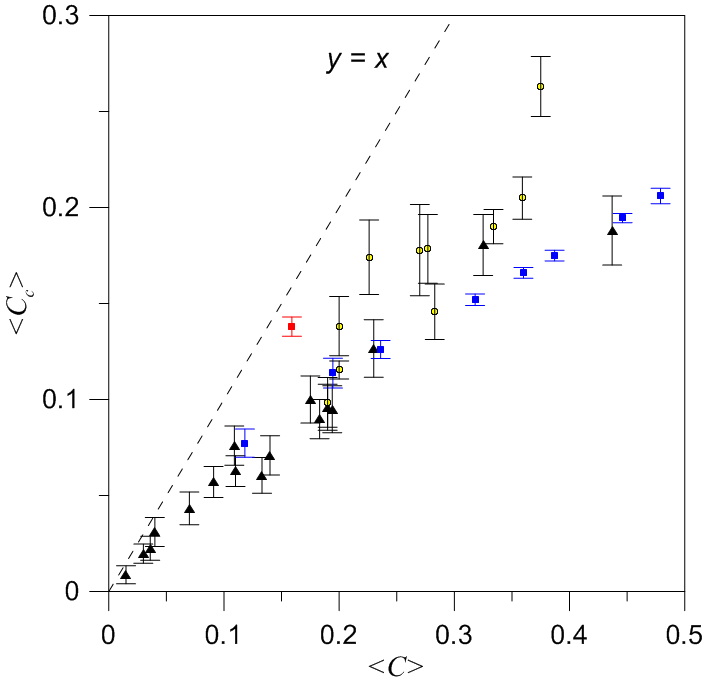}
		\includegraphics [width=0.45\textwidth] {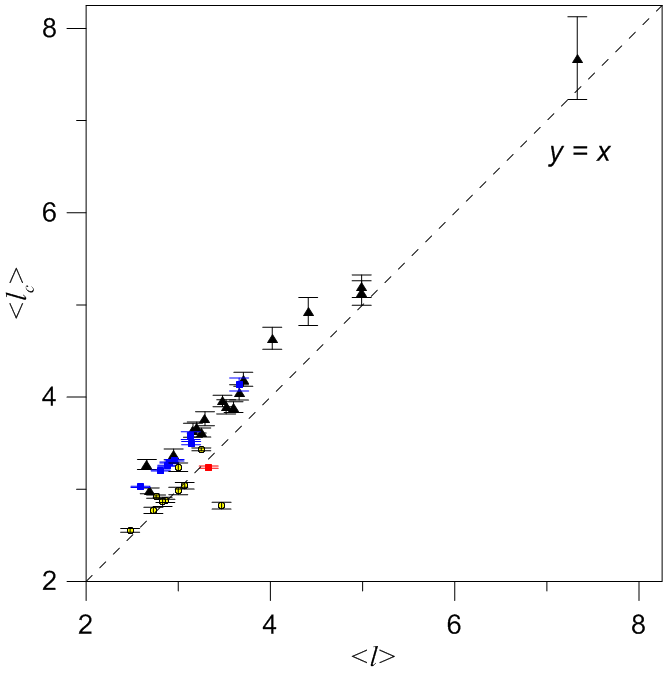}
		\caption{Mean clustering  $\langle C \rangle$  (Top) and average path length $\langle l \rangle$ (Bottom). Comparison of the original texts and their ciphered versions (\textit{c}). \textit{The Universal Declaration of Human Rights} (black triangles),  classic books (blue squares), computer codes (yellow squares) and ``\textit{Voynich manuscript}'' (red square).}
	\label{fig:L-C}
\end{figure} 

The top plot of Figure \ref{fig:L-C} shows that for most of the studied networks, except for the ``\textit{Voynich manuscript}'' and programming codes (red square and yellow circles, respectively), there is a much higher mean clustering coefficient than would be expected given only their ciphered versions denoting high transitivity of word connections in the original word networks. However, their average path length values are practically the same as the ones observed in the ciphered texts (Fig. \ref{fig:L-C}, bottom plot). 

The clustering coefficient, then, would seem to be a possible first step in solving the sensical text classification problem.  At the very least, focusing on the clustering coefficient might allow one to find differences between an original text and its ciphered version; however, this is both insufficient and misleading.  As Table \ref{tab:TABLEHR} shows, the clustering coefficient varies widely among the same text written in different languages. For example, the mean clustering of \textit{The Universal Declaration of Human Rights}, written in German, Spanish, or English, is on the order of the ciphered text written in Mahori (0.18); or the Basque text to the ciphered Russian Declaration (0.04). However an interesting property was found in this study: a correlation between the mean clustering  coefficient  and the average path length that follows a potential function, $\langle C \rangle \sim \langle l \rangle^{-3.56}$.

This property, far from trivial, seems to be exclusively a property of texts, differing from correlations found in other complex networks.

\begin{figure}[ht]
	\centering
		\includegraphics [width=0.5\textwidth] {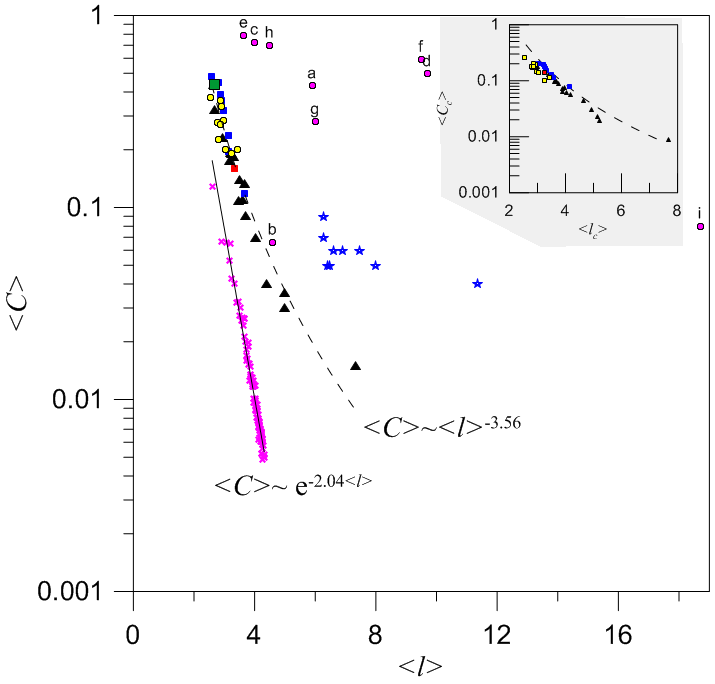}
\caption{Mean clustering  coefficient  $\langle C \rangle$  and average path length $\langle l \rangle$ correlation for different networks: word networks (color code as in Fig. \ref{fig:L-C}), social networks (a, b, c, d \cite{newman1,newman2,newman3}; e \cite{WS}, f, g \cite{barabasi2}), language network (h \cite{albert}), technological networks (blue stars \cite{sdh1,sdh2} and i \cite{WS}), another network of co-ocurrent words (green square \cite{albert}) and different networks generated by Barab\'asi model (pink crosses).}
	\label{fig:c_vs_l}
\end{figure} 

In Figure \ref{fig:c_vs_l} we can see that all the networks of co-ocurrent words follow the $\langle C \rangle \sim \langle l \rangle^{-3.56}$ potential function, and their position on this fit depends on their mean degree $\langle k \rangle$  (see Table \ref{tab:TABLEHR}). Thus, texts with high $\langle k \rangle$, are those with high clustering and low average path length. The text with the lowest mean degree, the Zulu Declaration of Human Rights, is the one furthest to the right of the curve ($\langle C \rangle$ = 0.015 and $\langle l \rangle$ = 7.33). It is interesting to note that when these texts are ciphered, they follow the same potential fit (see inset of the figure), but shift their positions to the right of the curve. This may be due to the way in which we made the word permutation, but it does suggest that nonsensical texts can also satisfy this relation. It is also interesting to highlight that both formal programming languages (yellow circles) and the ``\textit{Voynich manuscript}'' (red square) also fit this function. This function seems to be exclusive to texts, in fact, other complex networks of a different nature in the figure (social networks: a, c and d \cite{newman1,newman2,newman3}; e \cite{WS}; f and g \cite{barabasi2}), language networks: h \cite{albert} and technological networks: blue stars \cite{sdh1,sdh2}, i \cite{WS}) do not display this behavior, save network b  \cite{newman1,newman2,newman3}. Moreover, it is noteworthy that networks generated by Preferential Attachment (PA) (Barab\'asi network \cite{barabasi}) clearly show a different behavior (purple x) when compared to word networks. This result suggests that a PA-like mechanism is not valid for text construction, as \cite{Simon} otherwise suggests. In that model, increasing the number of nodes (and therefore the links between ``words'')  shifts the position in the plane  $\langle C \rangle vs \langle L \rangle$ to the right (higher values of $\langle l \rangle$, and lower of $\langle C \rangle$). This is the opposite of what happens in the case of word networks. These networks generally move to the right in the graph when the number of links decreases (actually when  $\langle k \rangle$ decreases, but those texts with smaller $E$ are the same as those with lowest mean degree $\langle k \rangle$, as shown in the Table \ref{tab:TABLEHR}.)

We have therefore found a property that can help discern whether a complex network comes from a text or not. In other words, in the presence of a network generated by linking words from a text based on co-ocurrence, that network should necessarily be positioned on the fit $\langle C \rangle \sim \langle L \rangle^{-3.56}$. However, this method still does not solve the deeper problem that motivates this work: the search for properties of networks of sensical texts. Ciphered texts also adjust to the function, not to mention texts written for machines and the ``\textit{Voynich manuscript}''. Correlation between clustering and average path length is a necessary condition, but still insufficient.

Notwithstanding the inability of that metric alone in describing sensical texts, we found another property that can help us to solve the puzzle: the network degree-assortativity \cite{Asso}. In the scenario of symmetric connections (undirected network) like the ones studied in this work, if densely connected nodes are connected to other nodes with many connections, then the network is considered assortative, $r>0$. On the other hand, if densely connected nodes are connected with their poorly connected counterparts (or \textit{vice versa}), then the network is disassortative, $r<0$. If no degree correlation is observed, $r\sim0$, then there is no link preference between nodes, as is the case in random networks or networks generated by the Barab\'asi model \cite{Asso}.

\begin{figure}[h]
	\centering
		\includegraphics [width=0.45\textwidth] {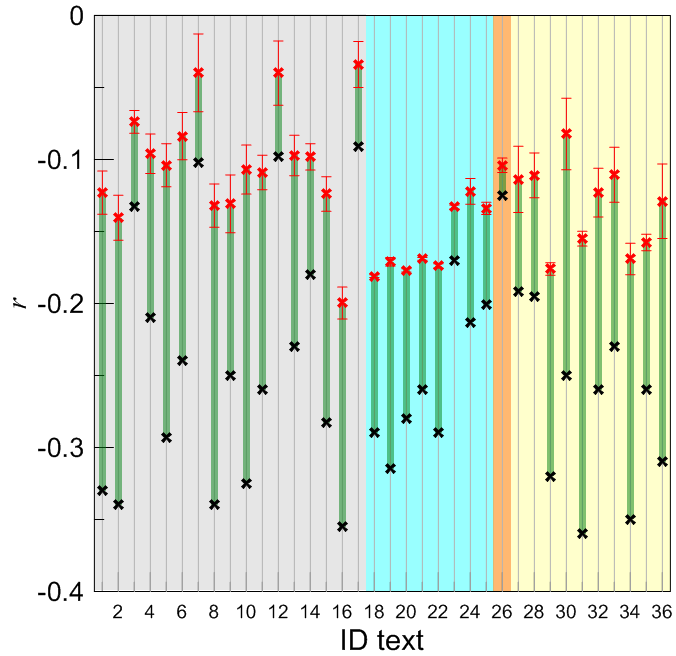}
		\includegraphics [width=0.45\textwidth] {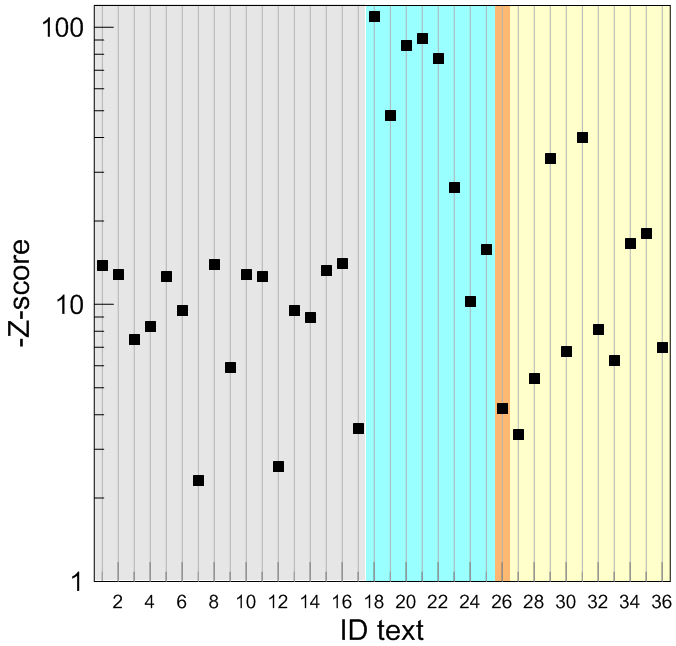}
		\caption{Top: Text (ID) \textit{vs} Assortativity $r$. Black crosses represent original texts, red crosses ciphered texts and green bars the difference. Bottom: Text (ID) negative Z-score Assortativity. Color regions: \textit{The Universal Declaration of Human Rights} (grey), classic books (blue),  ``\textit{Voynich manuscript}'' (orange), computer codes (yellow). }
	\label{fig:r}
\end{figure}

Word networks are expected to be disassortative  \cite{foster}, since highly connected words, such as articles, are linked with others that appear far less in the text, like nouns. It is necessary to emphasize that network b in Figure \ref{fig:c_vs_l} and other social networks, are typically assortative \cite{Asso}. The upper plot of Figure \ref{fig:r} shows the disassortative character of words networks.

In the figure, we can appreciate the assortativity values for word networks constructed from texts of Table \ref{tab:TABLEHR} (black crosses), and the mean assortativity of 100 networks constructed from the same texts but ciphered (red crosses) according to the same method of word permutation described above. As can be observed, original text network assortativities are negative (\textit{i.e.}, words networks are disassortative). However, the value for this correlation is much higher (and tends to 0) when the texts are ciphered. An interesting result appears, again, in the analysis of the ``\textit{Voynich manuscript}'' (ID=26). It is the only analyzed text that does not present a significant difference between assortativities before and after word permutation process. One book, ``\textit{Harry Potter and the Philosopher's Stone}'' (ID=23), also displays a smaller change. However, this text also shows other particular properties, possibly related to the target audience to which it was directed; in fact, this text had the highest value of mean degree (see table \ref{tab:TABLEHR}) and contains approximately 2000 words less than its version in Spanish (ID=22). To check the statistical validity of the change in the assortativity, we calculated the standard score (Z-score) \cite{krey} (bottom plot of Fig. \ref{fig:r}) according to,

\begin{equation}
Z_{score}= \frac{r(\text{ID}) - \langle r_c(\text{ID}) \rangle}{\sigma_c}
\end{equation}

\noindent where $r(\text{ID})$ is the assortativity of the text ID (see table \ref{tab:TABLEHR}),  $\langle r_c(\text{ID}) \rangle$  and $\sigma_c$ the mean and standard deviation of its ciphered versions, respectively.

\begin{figure}[h]
	\centering
		\label{fig:r2}
\end{figure} 

Significantly, and in all cases, the assortativity values of original text networks are below that of their ciphered versions, and particularly so for books (blue region). 

Although the word networks assortativity now allows us to discriminate between sensical texts (albeit for human or machine) and nonsensical (or ciphered) texts, it does not say anything by itself due to the wide range of values observed in all the sensical texts evaluated.

\begin{figure}[h]
	\centering
		\includegraphics [width=0.45\textwidth] {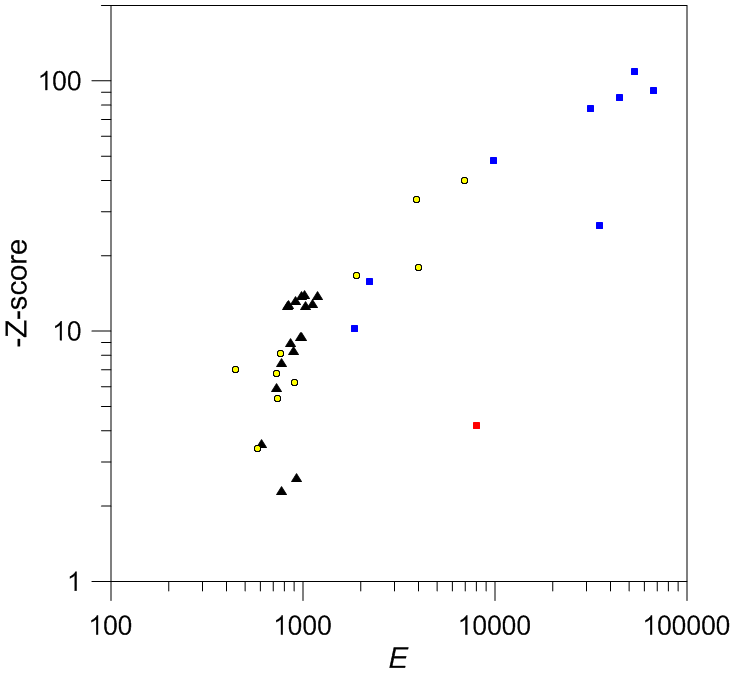}
		\caption{Z-score(Assortativity) and number of network edges $E$ correlation. Code color and shape of points as in Fig. \ref{fig:L-C}.}
	\label{fig:zs-e}
\end{figure} 

The different Z-score ranges shown in bottom plot of Figure \ref{fig:r} are explained by the fact that the texts have different lengths: the statistical models begin to fail when the text has fewer edges, since there are not as many possible variations. However, when a text has many edges, each successive realization in the process described herein gives a very different network, and eventually allows for significant observations of how far a corpus is from a sensical text value.  Confirming this, there is a positive correlation between Z-score and number of edges, as shown in Figure \ref{fig:zs-e}. 

\begin{figure}[!h]
	\centering
		\includegraphics [width=0.45\textwidth] {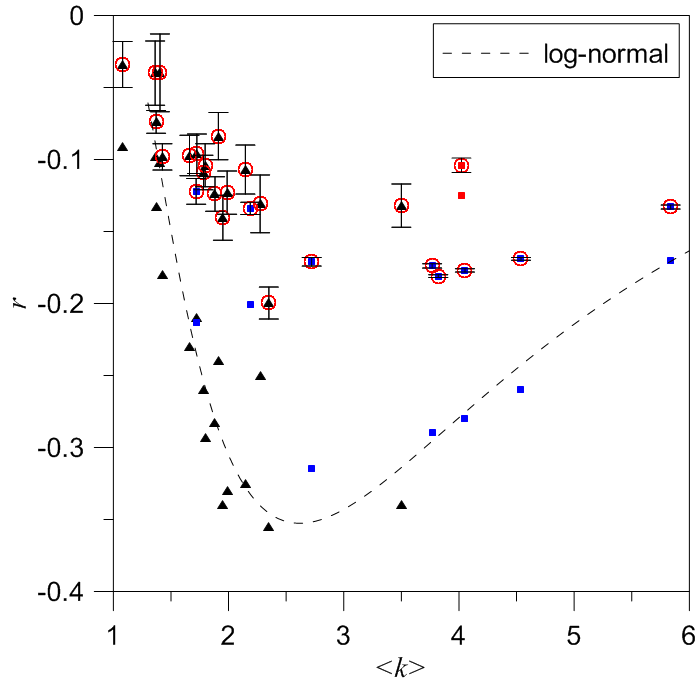}
		\caption{Degree assortativity $r$ and mean degree $\langle k \rangle$ correlation for texts ID=[1,26]. Code color and shape of points as in Fig. \ref{fig:L-C}. Ciphered texts marked with red circle.}
	\label{fig:formal}
\end{figure} 

In order to solve this problem with classifying texts as sensical using only assortativity values, we found another correlation that neither a ciphered text nor texts written in formal languages seem to follow.  The correlation between assortativity and mean degree of sensical texts written in natural languages follows a log-normal function, while any ciphered version moves away from this function (Fig. \ref{fig:formal}). This function permits us, then, to find a specific assortativity value for a certain mean degree for which a text is more likely to make sense. In fact, the text ID=23, which has the highest mean degree, also fits the function, while the ``\textit{Voynich manuscript}'' does not. The anomalous position of the manuscript in Figures \ref{fig:zs-e} and  \ref{fig:formal}, taken with its anomalous behaviors with respect to clustering and assortativity between its original and ciphered versions, suggest that the manuscript is a ciphered text, even with a lower disassortativity. We use the term ciphered because the manuscript has a word frequency distribution that follows Zipf's law, and because the correlation between $\langle C \rangle$ and $\langle l \rangle$ also fits the power function of other texts (see Fig. \ref{fig:c_vs_l}).

\begin{figure}[!h]
	\centering
		\includegraphics [width=0.45\textwidth] {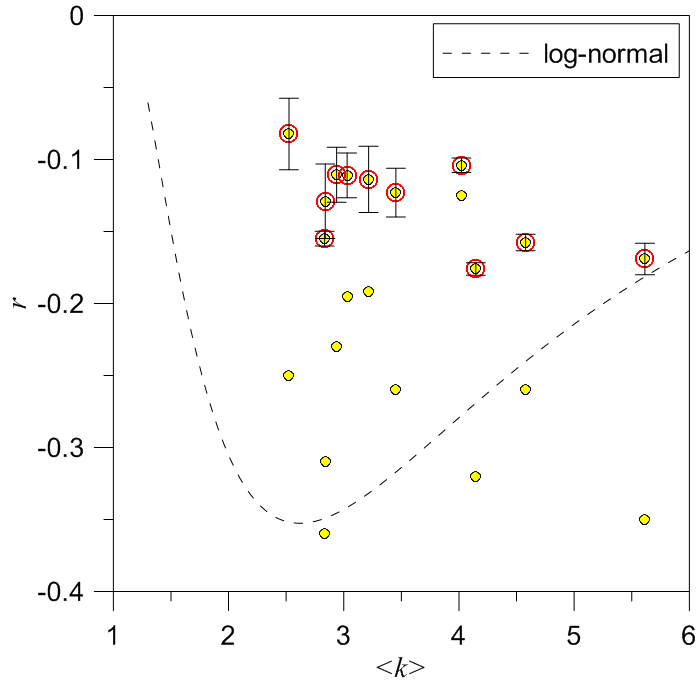}
		\caption{Degree assortativity $r$ and mean degree $\langle k \rangle$ correlation for computer codes, texts ID=[27,36]. Color code as in Fig. \ref{fig:formal}.}
	\label{fig:formal2}
\end{figure}

Finally, this log-normal function that fits the correlation between assortativity $r$ and mean degree $\langle k \rangle$, as well as Zipf's law, seems to be exclusive to texts written in natural languages. In fact, all the computer codes included in this study are outside this regularity, and even more so if they were ciphered (Fig. \ref{fig:formal2}).

\section{Conclusions}

In this work we searched for a method to discern between sensical texts, nonsensical/ciphered texts and texts written using formal grammar. Our approach was to analyze the topological properties of co-ocurrent word networks.

Our results suggest that a set of metrics related to network assortativity are able to solve this classification, as long as the word network has passed certain requirements. This means that if a network has been constructed from a sensical text, irrespective of its origin, we should expect that the network displays a ``proper'' position in the space $\langle C \rangle$ \textit{vs} $\langle l \rangle$, is disassortative, and that its correlation between assortativity and mean degree fits a log-normal function. However, if a word network has been constructed from a sensical text written in a formal programming language, then the last condition will most likely not be fulfilled.

This allows us to speculate on a desirable sorting method for the books in Library of Babel.  It might be possible to separate sensical texts from those that no one could understand, using the following method: (i) calculate four macroscopic statistical properties of word networks: mean degree, mean clustering coefficient, average path length, and network assortativity; (ii) check for the position of the word networks in plane $\langle C \rangle$ \textit{vs} $\langle l \rangle$. If the network does not fit the power function of Figure \ref{fig:c_vs_l}, it is a network that is not a text, and therefore should be discarded. If the text fits the function, the network has a text structure, but is not necessarily sensical, then; (iii) two alternatives are possible: (a) order these word networks by mean degree, and then from lowest to highest values of disassortativity. Those with higher values of disassortativity have a higher probability of being sensical texts. (b) Check for the position of these networks in plane $r$ \textit{vs} $\langle k \rangle$ of Figure \ref{fig:formal}. Those networks that are closer to the log-normal function, are more readable (\textit{i.e.}, make more sense) than versions located away from the fit line. Networks located far away from the fit are texts that might correspond to a ciphered text or texts written for a machine.

The result of our study leads us to propose the following hypothesis: for all variations of a text (with the same number and frequency of words), the most disassortative version will have the highest probability of making sense to a reader. Hence, this method could help in deciding when a deciphered version of a text is correct or not, without the necessity of ``translating'' the ciphered text. Simply put, the results from a deciphering process will be more effective than another if the resulting word network has a higher disassortativity.

Finally, our study has also allowed us to find evidence supporting the thesis that the ``\textit{Voynich manuscript}'' is a written text which has been ciphered, possibly by a permutation process of its words.

This work opens up interesting questions that will be addressed in future works.


\newpage

\begin{thebibliography}{99}

\bibitem {albert}
R. Albert and A.-L. Barab\'asi. 2002. Statistical mechanics of complex networks. \emph{Reviews of modern Physics}, \textbf{74}, 47--97.

\bibitem {barabasi}
A.-L. Barab\'asi and R. Albert. 1999. Emergence of scaling in random networks. \emph{Science}, \textbf{286}, 509--512.

\bibitem {barabasi2}
A.-L. Barab\'asi, H. Jeong, E. Ravasz, Z. Neda, A. Schubert, and T. Vicsek. 2001. preprint cond-mat/0104162.

\bibitem {BAb}
A.-L. Barab\'asi. 2009. Scale-Free Networks: A Decade and Beyond. \emph{Science}, \textbf{325}. 

\bibitem {borges}
J.L. Borges. 1981. \emph{El jard\'in de los senderos que se bifurcan}, Ficciones. Editorial Alianza. 


\bibitem {cam} 
J.P., C\'ardenas, M. Mouronte, J. C. Losada and R.M. Benito. 2010. Compatibility as underlying mechanism behind the evolution of networks. \emph{Physica A}, \textbf{389}, 1789--1798.

\bibitem {sdh1} 
J.P. C\'ardenas, A. Santiago, M.L. Mouronte, V. Feliu and R.M. Benito. 2010. Topological Analysis of Complex Optical Transport Networks. \emph{Int. J. Bifurcation and Chaos}, \textbf{20}(3), 787--794.

\bibitem {sdh2} 
J.P. C\'ardenas, M.L. Mouronte, L.G. Moyano, M.L. Vargas and R.M. Benito. 2010. On the Robustness of Spanish Telecommunication Networks. \emph{Physica A}, \textbf{389}(19), 4209--4216. 

\bibitem{Dor}
S.N. Dorogovtsev and J.F.F. Mendes. 2003. \textit{Evolution of Networks. From Biological Nets to the Internet and WWW}. Oxford University Press, New York, Oxford.

\bibitem {Erd}
P. Erd\H{o}s and A. R�nyi. 1959. On random graphs. \emph{Publ. Math. Debrecen} \textbf{6}, 290--297.

\bibitem {Fer1}
R. Ferrer i Cancho and R.V. Sol\'e. 2001. The Small-World of Human Language. \emph{Proceedings of the Royal Societyof London B}, \textbf{286}: 2261-2266.


\bibitem {Fer3}
Ferrer-i-Cancho R. and Elvevag B. 2010. Random texts do not exhibit the real Zipf's law-like rank distribution. \emph{PloS ONE}, \textbf{5}: e9411. 

\bibitem{foster}
J.G. Foster, D.V. Foster, P. Grassberger and Maya Paczuski. 2010. Edge direction and the structure of networks. \emph{PNAS}, \textbf{107}(24), 10815--10820.

\bibitem {krey}
E. Kreyszig. 1979. \emph{Applied Mathematics}, Wiley Press.

\bibitem {landini}
G. Landini. 2001. Evidence of linguistic structure in the Voynich Manuscript using spectral analysis. \emph{Cryptologia}, \textbf{25}(4), 275--295.

\bibitem{lui}
H. Lui and J. Cong. 2012. Language clustering with word co-occurrence networks based on parallel texts. \emph{Chinese Science Bulletin}, \textbf{58}(10), 1139--1144.

\bibitem {montemurro1}
M.A. Montemurro, D.H. Zanette. 2010. Towards the quantification of the semantic information in written language. \emph{Advances in Complex Systems}, \textbf{13}, 135--153.

\bibitem {montemurro}
M.A. Montemurro, D.H. Zanette. 2010. Keywords and Co-Occurrence Patterns in the Voynich Manuscript: An Information-Theoretic Analysis. \emph{PLoS ONE}, \textbf{8}(6): e66344. 

\bibitem {newman1}
M. E. J. Newman. 2001a. The structure of scientific collaboration networks. \emph{PNAS}, \textbf{98}(2), 404--409.

\bibitem {newman2}
M. E. J. Newman. 2001b. Scientific collaboration networks: I. Network construction and fundamental results. \emph{Phys. Rev. E}, \textbf{64}, 016131.

\bibitem {newman3}
M. E. J. Newman. 2001c. Scientific collaboration networks: II. Shortest paths, weighted networks, and centrality. \emph{Phys. Rev. E}, \textbf{64}, 016132.

\bibitem {Asso}
M. E. J. Newman. 2002. Assortative Mixing in Networks. \emph{Physical Review E} \textbf{89}, 208701.

\bibitem {New}
M. E. J. Newman. 2003. The Structure and Functions of Complex Networks. \emph{SIAM Review}, \textbf{45}(2),167--256.


\bibitem {Newb}
M. E. J. Newman, A-L. Barab\'asi and D. Watts. 2006. \emph{The Structure and Dynamics of Networks}, Princeton University Press.

\bibitem {reghizzi}
S. C. Reghizzi. 2009. \emph{Formal Languages and Compilation}, Texts in Computer Science, Springer.

\bibitem {rugg}
G. Rugg. 2004. An elegant hoax? a possible solution to the Voynich manuscript. \emph{Cryptologia}, \textbf{28}, 31--46. 

\bibitem {schinner}
A. Schinner. 2007. The Voynich manuscript: Evidence of the hoax hypothesis. \emph{Cryptologia}, \textbf{31}, 95--107. 

\bibitem {Simon}
H. A. Simon. 1955. On a class of skew distribution functions, \emph{Biometrika} \textbf{42}, 425--440.

\bibitem {sole}
R.V. Sol\'e. 2009. \textit{Redes Complejas. Del genoma a Internet}. Editorial Tusquets.

\bibitem {WS}
D.J. Watts and S.H. Strogatz. 1998. Collective dynamics of Small-World networks. \emph{Nature} \textbf{393}, 440--442.

\bibitem {zipf}
G.L. Zipf, (1965), \textit{Human Behavior and the Principle of Least Effort}. Hafner, New York.

\end{thebibliography}
\end{document}